\documentclass[structabstract,letter]{aa}  
\usepackage{graphicx}
\usepackage{txfonts}
\usepackage{natbib}
\usepackage{breqn}
\usepackage{longtable,lscape}
\usepackage{hyperref}

\begin{document}

   \title{A reevaluation of the 2MASS zero points using \\ CALSPEC spectrophotometry complemented with \\ {\em Gaia} Data Release 2 parallaxes}


   \titlerunning{2MASS ZPs from CALSPEC and Gaia DR2}

   \author{J. Ma{\'\i}z Apell{\'a}niz\inst{1}
           \and
           M. Pantaleoni Gonz\'alez\inst{1,2}
          }

   \institute{Centro de Astrobiolog{\'\i}a, CSIC-INTA. Campus ESAC. Camino bajo del castillo s/n. E-28\,692 Villanueva de la Ca\~nada. Spain. \\
              \email{jmaiz@cab.inta-csic.es} \\
              \and
              Departamento de Astrof{\'\i}sica y F{\'\i}sica de la Atm\'osfera. Universidad Complutense de Madrid. E-28\,040 Madrid. Spain. \\
             }

   \authorrunning{Ma{\'\i}z Apell{\'a}niz \& Pantaleoni Gonz\'alez}

   \date{Received XX XXX 2018; accepted XX XXX 2018}

 
  \abstract
  {2MASS is the reference survey in the NIR part of the spectrum given its whole-sky coverage, large dynamic range, and proven calibration uniformity. However, 
   previous studies disagree in the value of the zero points (ZPs) for its three bands $JHK$ at the hundredth of a magnitude level. The disagreement should become
   more noticeable now that Gaia provides whole-sky optical photometry calibrated below that level.}
  {We want to establish the value of the 2MASS ZPs based on NICMOS/HST spectrophotometry of the CALSPEC standard stars and test it with the help of Gaia DR2 parallaxes.}
  {We have computed the synthetic $JHK$ photometry for a sample of stars using the HST CALSPEC spectroscopic standards and compared it with their 2MASS magnitudes to
   evaluate the ZPs. We have tested our results by analysing a sample of FGK dwarfs with excellent 2MASS photometry and accurate Gaia DR2 parallaxes.}
  {The Vega ZPs for 2MASS $J$, $H$, and $K$ are found to be $-0.025\pm 0.005$ mag, $0.004\pm 0.005$ mag, and $-0.015\pm 0.005$ mag, respectively. The analysis of FGK sample indicates that 
  the new ZPs are more accurate than previous ones.}
  {}


   \maketitle
%

\section{Introduction}

$\,\!$ \indent The precise calibration of photometric surveys requires an almost constant reevaluation, as more data are accumulated over time and new surveys introduce new 
opportunities to test their compatibility. The Two Micron All Sky Survey (2MASS, \citealt{Skruetal06}) revolutionized infrared astronomy in the previous decade, as it introduced what is 
now the golden standard in NIR photometry in terms of sky coverage, dynamic range, and stability for its three $JHK$ bands. The recent {\it Gaia} Data Release 2 (DR2, \citealt{Browetal18}) 
is likely to produce a similar revolution in optical photometry with its whole-sky $GG_{\rm BP}G_{\rm RP}$ photometry that will be expanded into spectrophotometric information in future data 
releases.

In this letter we calculate new values for the 2MASS $JHK$ Vega zero points (from now on, ZPs for notation simplicity) using the recently updated CALSPEC library 
\citep{Bohletal17}\footnote{See also \url{http://www.stsci.edu/hst/observatory/crds/calspec.html}.} in section 2. In section 3 we check them using FGK dwarfs with Gaia DR2 parallaxes.
For consistency with our previous work, we use the Vega spectral energy distribution (SED) provided by 
\citet{Bohl07}\footnote{Available from \url{ftp://ftp.stsci.edu/cdbs/calspec/alpha_lyr_stis_003.fits}.} 
and the 2MASS sensitivity functions provided by the 2MASS web site\footnote{\url{https://www.ipac.caltech.edu/2mass/releases/allsky/doc/sec6_4a.html\#rsr}.}. The reader can use the appendix
if interested in converting the results to other magnitude systems.

\section{New 2MASS Vega zero points from CALSPEC}

\begin{figure}
\centerline{\includegraphics[width=0.98\linewidth, bb=28 48 566 506]{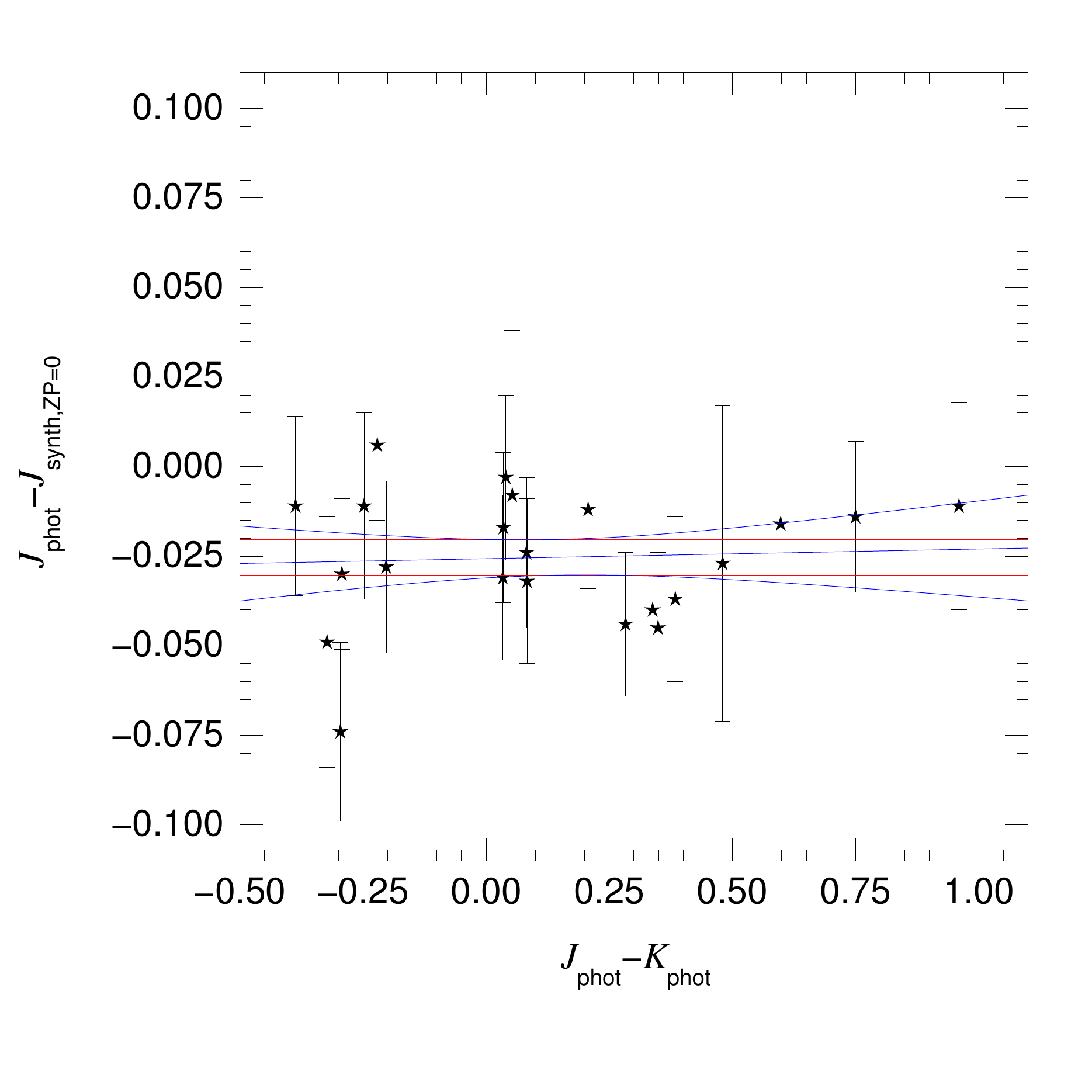}}
\centerline{\includegraphics[width=0.98\linewidth, bb=28 48 566 506]{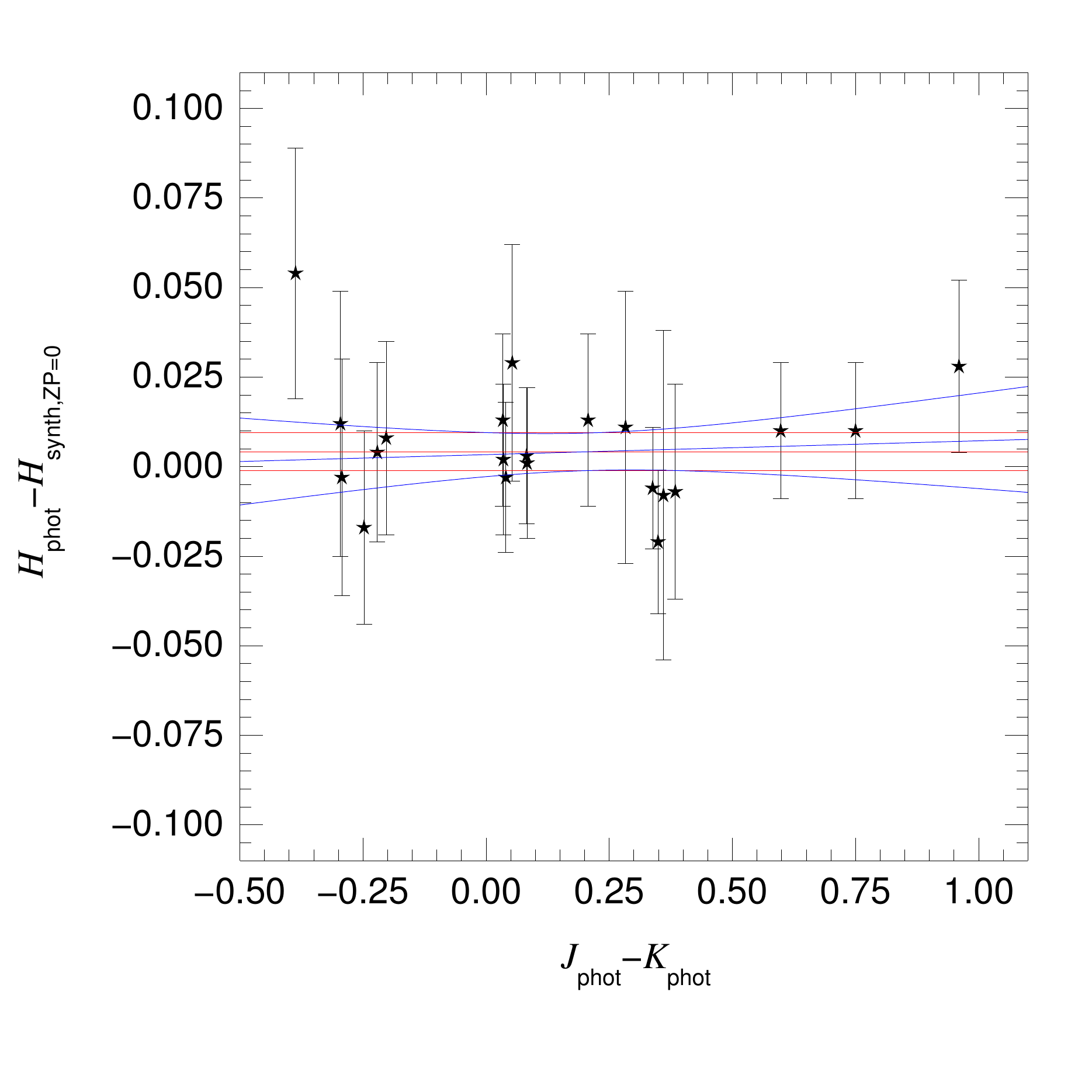}}
\centerline{\includegraphics[width=0.98\linewidth, bb=28 48 566 506]{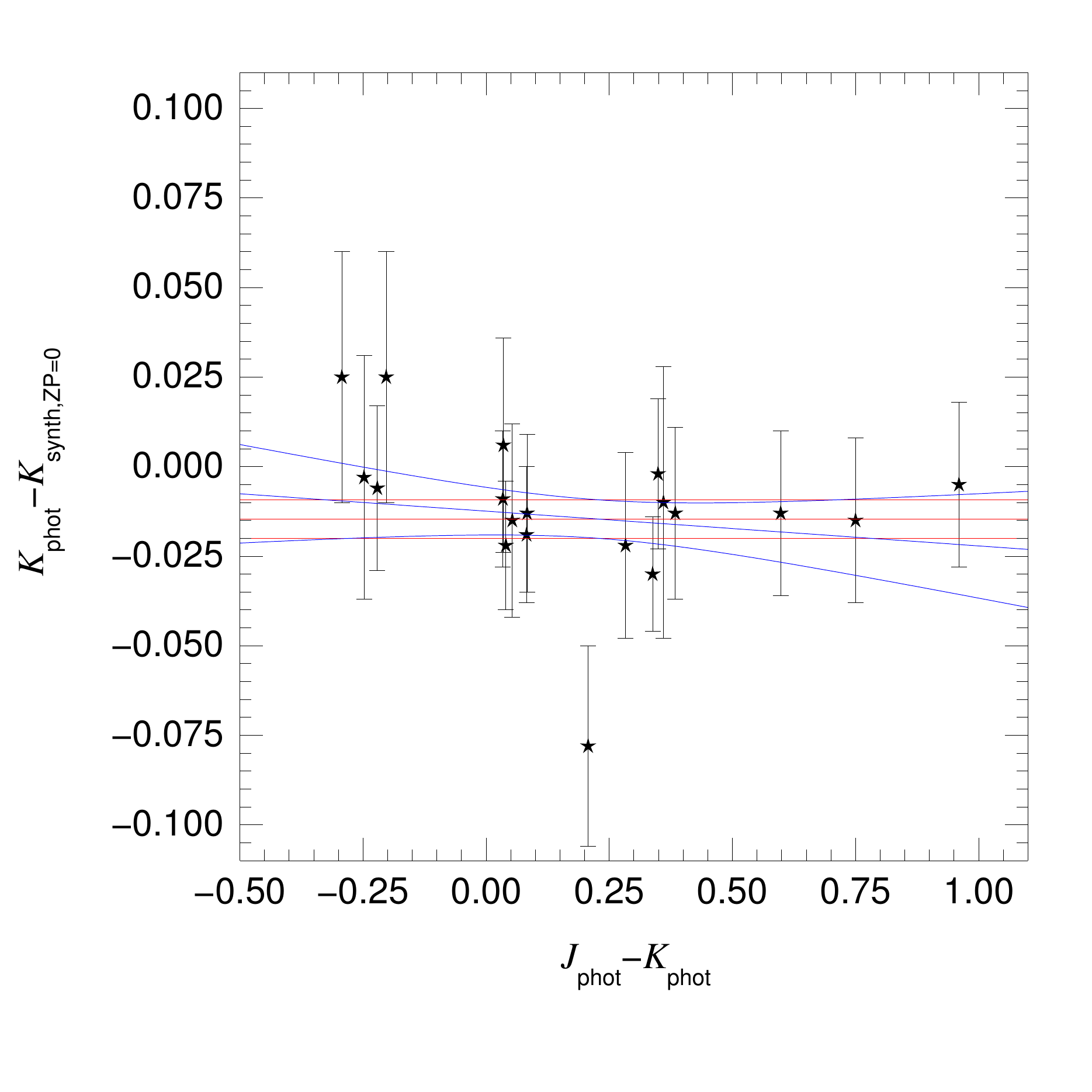}}
\caption{Difference between the photometric and the synthetic magnitudes as a function of photometric $J-K$ for the CALSPEC stars in this sample assuming a Vega ZP of 0.0 for 2MASS 
         $J$ (top), $H$ (center), and $K$ (bottom). For each plot we show the fit plus uncertainties assuming a linear color term (blue) and no color term (red).}
\label{photsynth}
\end{figure}

$\,\!$ \indent The technique we apply in this paper has been used in the past \citep{Maiz05b,Maiz06a,Maiz07a,Maiz17} to test different photometric systems. We compute the synthetic
photometry of stars observed with high-quality HST spectrophotometry and we compare it with the observed photometry. If the assumed sensitivity function corresponds to the actual one, a
plot of the difference between the observed and synthetic magnitudes (or colors) as a function of color should be essentially flat (i.e. have no color terms). The weighted mean of the 
values in the vertical axis is the ZP and the standard deviation of their mean its uncertainty. If, on the other hand, a color term is present, one needs to go back to the drawing board and
derive a new sensitivity function \citep{Weiletal18}. 

\begin{table}
\caption{CALSPEC sample used for each filter.}
\centerline{
\begin{tabular}{lll}
\hline
2MASS ID & CALSPEC file & Filters \\
\hline
J05053062$+$5249519 & \tt{g191b2b\_stisnic\_006.fits}      & $JHK$ \\
J05522761$+$1553137 & \tt{gd71\_stisnic\_006.fits}         & $JH$  \\
J09211915$+$8143274 & \tt{agk\_81d266\_stisnic\_006.fits}  & $JHK$ \\
J12570233$+$2201526 & \tt{gd153\_stisnic\_006.fits}        & $JH$  \\
J13385054$+$7017077 & \tt{grw\_70d5824\_stisnic\_007.fits} & $JHK$ \\
J14515797$+$7143173 & \tt{p041c\_stisnic\_007.fits}        & $JHK$ \\
J15591357$+$4736419 & \tt{p177d\_stisnic\_007.fits}        & $JHK$ \\
J16194609$+$5534178 & \tt{snap2\_stisnic\_007.fits}        & $J$   \\
J16313382$+$3008465 & \tt{p330e\_stisnic\_008.fits}        & $JHK$ \\
J16553529$-$0823401 & \tt{vb8\_stisnic\_006.fits}          & $JHK$ \\
J17325264$+$7104431 & \tt{1732526\_stisnic\_004.fits}      & $JHK$ \\
J17403468$+$6527148 & \tt{1740346\_stisnic\_003.fits}      & $JHK$ \\
J17430448$+$6655015 & \tt{1743045\_stisnic\_004.fits}      & $JHK$ \\
J17551622$+$6610116 & \tt{kf08t3\_stisnic\_001.fits}       & $JHK$ \\
J17583798$+$6646522 & \tt{kf06t2\_stisnic\_004.fits}       & $JHK$ \\
J18022716$+$6043356 & \tt{1802271\_stisnic\_004.fits}      & $JHK$ \\
J18023073$+$5837381 & \tt{hd165459\_stisnic\_004.fits}     & $JHK$ \\
J18052927$+$6427520 & \tt{1805292\_stisnic\_004.fits}      & $JHK$ \\
J18120957$+$6329423 & \tt{1812095\_stisnic\_004.fits}      & $JHK$ \\
J21321623$+$0015144 & \tt{lds749b\_stisnic\_006.fits}      & $J$   \\
J22031077$+$1853036 & \tt{hd209458\_stisnic\_007.fits}     & $JHK$ \\
J22113136$+$1805341 & \tt{bd\_17d4708\_stisnic\_006.fits}  & $HK$  \\
J23195840$-$0509561 & \tt{feige110\_stisnic\_006.fits}     & $JHK$ \\
\hline
\end{tabular}
}
\label{nicmos}
\end{table}

We selected from the CALSPEC database those objects with NICMOS spectrophotometry \citep{BohlCohe08} and 2MASS photometric uncertainties lower than 0.05 mag. Those criteria leave 22, 21,
and 19 stars for $J$, $H$, and $K$, respectively (Table~\ref{nicmos}). The comparison between the observed and synthetic photometry is shown in Fig.~\ref{photsynth}.

For all three filters we detect no significant color term, as the restricted fit (no color term allowed) is always encompassed by the unrestricted fit in the plotted color range. Therefore, the
published sentitivity functions agree with the data presented here and no new functions need to be calculated. For the restricted fit, we obtain reduced $\chi^2$ values of 0.76, 0.59, and
0.75 for $J$, $H$, and $K$, respectively. Such low values indicate that if anything, the 2MASS uncertainties are slightly overestimated on average for these objects (note that we have
not included an uncertainty term from the spectrophotometry, but that would only make the reduced $\chi^2$ values even lower). 

The values for the ZPs derived from the restricted fit are given in Table~\ref{ZPs}, along with the equivalent results from previous works. The associated uncertainties for the ZPs obtained
here are 0.005 mag in all three cases. The ZP derived for $H$ is consistent with the previous values. The ZPs derived for $J$ and $K$ are both close to the lower end of previous results,
with the one for $J$ closer to our previous result (\citealt{Maiz07a}, which were obtained using a preliminary version of this analysis with a previous version of CALSPEC data) and
the one for $K$ closer to the result of \citet{Coheetal03}.

\begin{table}
\caption{ZPs based on the \citet{Bohl07} Vega SED derived in this work and in previous ones.}
\label{ZPs}
\centerline{
\begin{tabular}{lccc}
\hline
Reference          & $J$      & $H$      & $K$      \\
\hline
\citet{Coheetal03} & $-$0.006 & $+$0.007 & $-$0.022 \\
\citet{HolbBerg06} & $-$0.014 & $+$0.004 & $+$0.005 \\
\citet{Maiz07a}    & $-$0.021 & $+$0.009 & $+$0.000 \\
This work          & $-$0.025 & $+$0.004 & $-$0.015 \\
\hline
\end{tabular}
}
\end{table}

\section{Testing the zero points with FGK dwarfs}

$\,\!$ \indent We test our values for the 2MASS ZPs using additional information from Gaia DR2, most importantly the parallaxes $\varpi$. We start by cross-matching the whole 
2MASS catalog with Gaia DR2 and selecting only the objects with good-quality photometry in all six bands $GG_{\rm BP}G_{\rm RP}JHK$. 
We then select a sample of low-extinction FGK dwarfs by applying the following restrictions to the sample above:

\begin{itemize}
 \item $0.0 < J-H < 0.6$ and $-0.1 < H-K < 0.2$, which is the region where low-extinction FGK dwarfs are located \citep{Finletal00}.
 \item $\varpi/\sigma_\varpi > 10$ i.e. only stars with low relative-uncertainty parallaxes in order to be able to use the approximation for the distance $d = 1/\varpi$ without introducing 
       significant biases \citep{Maiz01a,Maiz05c,Lurietal18}.
 \item A double cone in Galactic latitude $|b| > 60\degr$ to minimize extinction. We have used \citet{SchlFink11} and the \citet{Maizetal14a} extinction law with $R_{5495} = 3.1$ to
       calculate that the average $A_1$ (extinction at 1~$\mu$m) at large distances in that region is 0.024~mag, which becomes our expected value (within the uncertainties as to the exact 
       location of the dust along the line of sight). 
 \item $\sigma_J < 0.025$ mag, $\sigma_H < 0.025$ mag, $\sigma_K < 0.025$ mag and 2MASS quality flag AAA to minimize noise.
 \item 100 pc $ < d < $ 1000 pc to avoid biases caused by saturation of bright stars and loss of dim stars.
 \item A cut in $K$ absolute magnitudes ($K - 5\log\,d + 5$) uncorrected for extinction to exclude subdwarfs and evolved stars.
\end{itemize}

\begin{figure}
\centerline{\includegraphics[width=\linewidth]{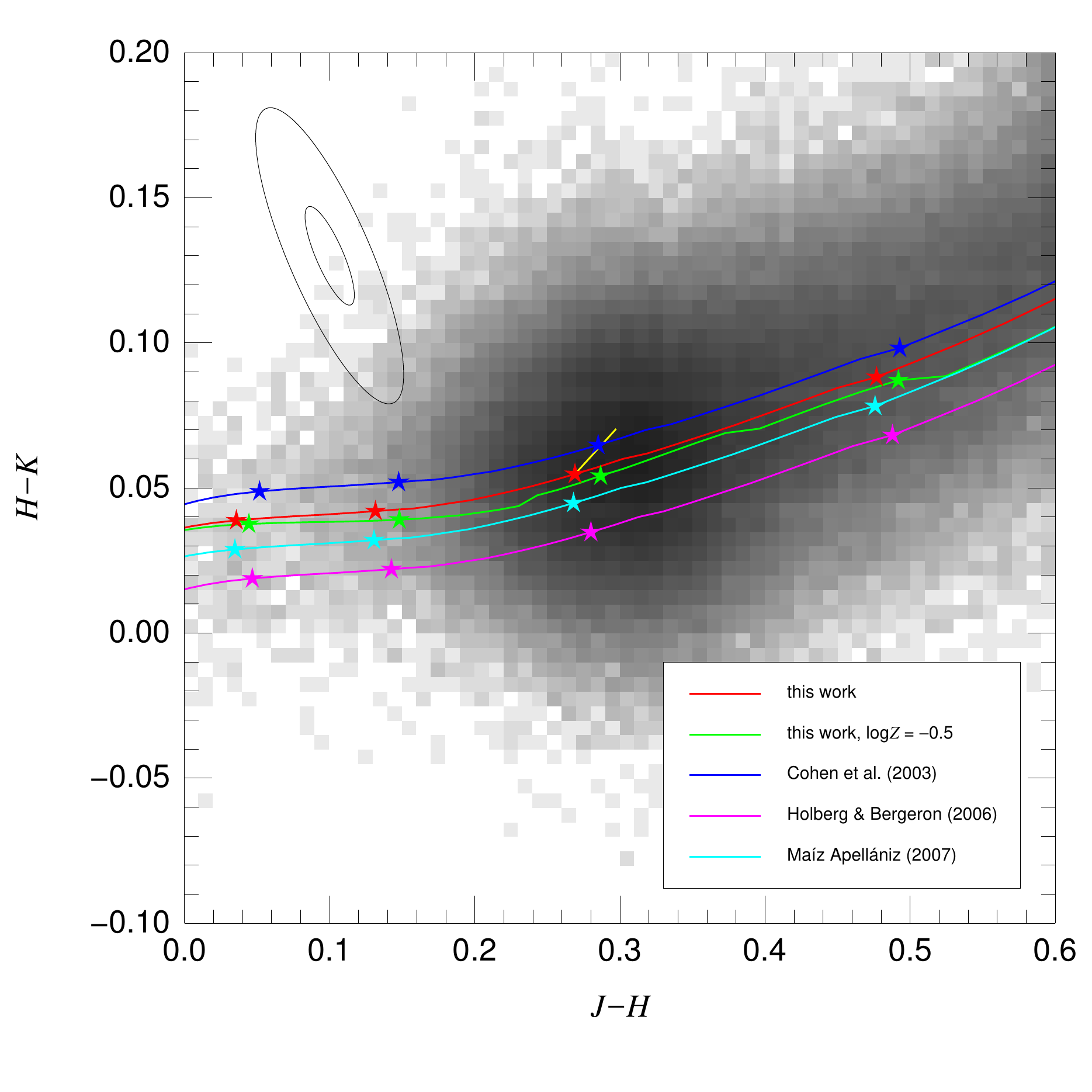}}
\caption{$J-H$ + $H-K$ density plot for the FGK dwarf sample described in the text. The intensity scale is logarithmic. The colored lines show the location of the \citet{BrotHaus05} dwarf 
models for four different sets of 2MASS ZPs assuming solar metallicity and for the ZPs in this work assuming $\log Z = -0.5$ in solar units. The colored stars mark the location of the models 
with $T_{\rm eff} = $ 8, 7, 6, and 5~kK (from left to right), respectively. The large ellipse shows the 68.5\% contour for a distribution assuming 
$\sigma_J = \sigma_H = \sigma_K = 0.0015$~mag and the small ellipse the equivalent for the ZP uncertainties in this work. The straight yellow line shows the displacement that corresponds to
an extinction with $A_1 = 0.1$ mag and $\alpha = 2.0$.}
\label{FGK}
\end{figure}

With those conditions, we are left with a sample of 216\,393 stars, whose density plot in the $J-H$ + $H-K$ plane is shown in Fig.~\ref{FGK}. We also plot there the location of the 
colors predicted by the \citet{BrotHaus05} SED models with solar metallicity assuming no extinction and the different Vega ZPs in Table~\ref{ZPs}. To explore the (small) effect of metallicity 
on intrinsic colors we show the case of $\log Z = -0.5$ (solar units) for the ZPs in this work. 

Most of the dispersion in Fig.~\ref{FGK} is caused by the photometric uncertainties (see large ellipse for a typical case); if they were smaller, we should see a much narrower distribution in 
the vertical direction. The effect of extinction is present but is much smaller: the yellow line shows the extinction trajectory for $A_1 = 0.1$ mag assuming a power-law NIR extinction 
$A_\lambda = A_1 \lambda^{-\alpha}$ with $\alpha = 2.0$. Note that is about four times the expected effect of $A_1 = 0.024$~mag for the sample, which for this extinction law corresponds to
$E(H-K) = 0.004$ mag\footnote{Strictly speaking, the value of $E(H-K)$ depends on the SED, as it is a filter-integrated quantity, but the differences are minimal in the NIR (much smaller than
1~millimagnitude) for the stars considered here \citep{Maiz13b,MaizBarb18}.}.

\begin{table}
\caption{Results for the FGK dwarf sample.}
\label{FGKresults}
\centerline{
\begin{tabular}{lccc}
\hline
ZPs                         & $<A_1>$  & $\sigma_{A_1}$ \\
\hline
\citet{Coheetal03}          & $-$0.055 & 0.257          \\
\citet{HolbBerg06}          & $+$0.203 & 0.243          \\
\citet{Maiz07a}             & $+$0.101 & 0.252          \\
This work                   & $+$0.012 & 0.258          \\
This work ($\log Z = -0.5$) & $+$0.049 & 0.250          \\
\hline
\end{tabular}
}
\end{table}

For each target we calculate $A_1$ by tracing back the extinction trajectory to the intrinsic color curve derived from the \citet{BrotHaus05} models
and we repeat it for the different ZPs in Table~\ref{ZPs}. Given that photometric
uncertainties are the most important cause for dispersion in Fig.~\ref{FGK}, many objects will have negative measured values of $A_1$. The results for the average $A_1$ are given in
Table~\ref{FGKresults}. The ZPs in this work are those which are closer to the expected value of 0.024~mag. \cite{Coheetal03} yield an average negative value of $A_1$ (the blue curve is 
displaced upwards)
while \citet{HolbBerg06} and \citet{Maiz07a} yield average values of $A_1$ that are too large (their curves are displaced downwards). The expected value lies between the cases in this work with
solar and subsolar metallicity, as expected for lines of sight near the Galactic poles (switching from solar to subsolar metallicity increases $A_1$ by 0.037 mag). 
Indeed, if most of the extinction originates close to the Galactic Plane, the population at short distances should have an average higher metallicity than that at long distances. 
For our method, that translates into a measured $A_1$ that artificially decreases with distance (if one does not correct for metallicity) and
we indeed see that effect in our data. 

We conclude that our ZPs provide a better fit to the intrinsic NIR colors of FGK stars than previous works. Note, however, that differences
between different sets of ZPs are relatively small: the small ellipse in Fig.~\ref{FGK} shows that the ZPs of \citet{Maiz07a} are within 1 sigma (in the color-color plane) and those of the
other two works are not far away. Even though we claim our results are better than previous ones and should be preferredly used, there are no large fundamental differences between the four 
calibrations. The reasons why there is not a better possible calibration of the 2MASS at this stage are two: on the one hand, the number of CALSPEC data is limited in number and, alas, with
NICMOS long gone there will be no more high-quality HST spectroscopy in the whole 1.0-2.5~$\mu$m range. On the other hand, 2MASS photometry saturates at 4-5 magnitudes and brighter stars in the
survey have large uncertainties that make them useless for calibration purposes. With the overall interest of the astronomical community into going deeper and deeper in magnitude, we are
forgetting the brighter stars which are not only interesting by themselves but are also the best calibration targets given that their high fluxes yield large S/N in short exposure times for
most programs. The problem with bright stars in the NIR is that most detector setups saturate them even at the shortest exposure times so this is a case where we need very small telescopes. If
it were not for the differences in the access to technology between the optical and the NIR, an excellent amateur astronomy project would be an all-sky NIR photometric survey.

\begin{acknowledgements}
This work has made use of data from: (a) The ESA/NASA {\it Hubble Space Telescope}, obtained from the data archive at the Space Telescope Science Institute (STScI). STScI is operated by 
the Association of Universities for Research in Astronomy, Inc. under NASA contract NAS 5-26555. (b) The Two Micron All Sky Survey (2MASS), which is a joint project of the University of 
Massachusetts and the Infrared Processing and Analysis Center/California Institute of Technology, funded by the National Aeronautics and Space Administration (NASA) and the National 
Science Foundation (NSF). For both institutions the word ``National'' refers to the United States of America. (c) The European Space Agency (ESA) mission {\it Gaia} 
({\tt https://www.cosmos.esa.int/gaia}), processed by the {\it Gaia} Data Processing and Analysis Consortium (DPAC, {\tt https://www.cosmos.esa.int/web/gaia/dpac/consortium}). Funding for 
the DPAC has been provided by national institutions, in particular the institutions participating in the {\it Gaia} Multilateral Agreement. We acknowledge support from the Spanish 
Government Ministerio de Ciencia, Innovaci\'on y Universidades through grant AYA2016-75\,931-C2-2-P. M.P.G. acknowledges support from the ESAC Trainee program.
\end{acknowledgements}

\bibliographystyle{aa}
\bibliography{general}

\begin{appendix}

\section{Zero points and conversions between magnitude systems}

$\,\!$\indent In order to compare observed photometric magnitudes ($m_{{\rm phot},p}$) in a series of filter passbands (denoted here by the $p$ index) with spectral energy distributions (SEDs, denoted
here by the $s$ index) $f_{\lambda,s}(\lambda)$ one has to define a magnitude system $r$ and compute synthetic magnitudes ($m_{r,p}$) from the SEDs. For a photon-counting detector and a total-system 
dimensionless sensitivity function $P_p(\lambda)$
 
\begin{equation}
m_{r,p}[f_{\lambda,s}(\lambda)] = 
 -2.5\log_{10}\left(\frac{\int P_p(\lambda)f_{\lambda,s}(\lambda)\lambda\,d\lambda}
                         {\int P_p(\lambda)f_{\lambda,r}(\lambda)\lambda\,d\lambda}\right)
                       + {\rm ZP}_{r,p}.
\label{photon}
\end{equation} 

A magnitude system $r$ is defined by a reference SED $f_{\lambda,r}(\lambda)$ and a series of (relative) zero points ZP$_{r,p}$ for each filter. Most current magnitude systems use one of these
reference SEDs:

\begin{itemize}
 \item Vega (or VEGAMAG): a measured Vega SED $f_{\lambda,{\rm Vega}}$.
 \item ST: $f_{\lambda,{\rm ST}} = 3.63079\cdot 10^{-9}$  erg s$^{-1}$ cm$^{-2}$ \AA$^{-1}$ (constant).
 \item AB: $f_{\nu,{\rm AB}}     = 3.63079\cdot 10^{-20}$ erg s$^{-1}$ cm$^{-2}$ Hz$^{-1}$ (constant).
\end{itemize}
 
In principle, one can adjust the $f_{\lambda,r}(\lambda)$ in such a way that ZP$_{r,p}$ = 0.0 for all filters, which generates the default or strict system for that SED. In practice, one calibrates 
observed magnitudes a posteriori and this results in (unavoidable) zero points, no matter what reference SED one uses\footnote{An alternative to this technique is used in some instances for HST 
photometry, where the ZPs are forced to be zero and the changes are introduced in the photometric reduction pipeline to change the observed photometric magnitudes instead of the synthetic ones. The 
problem with this aproach is that it introduces a source of confusion, as the pipeline may deliver a given magnitude for an observation of a star today and a different one for the same observation
tomorrow.}. If the analysis is done correctly the most one can expect is to have ZPs close to zero. For example, SDSS magnitudes use an AB reference system and have ZPs measured in hundredths of a
magnitude \footnote{See \url{https://www.sdss.org/dr12/algorithms/fluxcal/}.}

The 2MASS magnitude system was defined using Vega as the reference SED and, indeed, the ZPs we measure in this work are close to zero, as expected. However, Vega-based magnitudes have two problems. The
first one is that the published Vega SEDs differ from one another, so any published work must clearly state which one is using and make it available. In this paper we do so in the
introduction. Strictly speaking, it does not matter much how accurate (i.e. closer to the real one) the Vega SED that one uses is, as a change in the denominator in Eqn.~\ref{photon} is simply offset 
by a change in ZP$_{r,p}$. The second problem is that why should one bother with requiring a Vega SED and not use a simpler ST or AB reference SED instead. From the point of view of designing a
magnitude system from scratch that is a valid issue and is the reason why more modern systems such as SDSS have taken that step. From the point of view of calibrating a preexisting photometric system,
following that strategy leads into ZPs that are not close to zero and are a possible source of confusion. For that reason, the ZPs in this paper are expressed using Vega as a reference SED. 
Nevertheless, a valid point is that the Vega SED used may not be available so it should be possible to work with ST or AB magnitudes even for older photometric systems. The solution to this problem is 
given by \citet{Maiz07a}: to convert from a Vega system to an ST system one should use Eqn.~4 there and to do the same to an AB system one should use Eqn.~5 there. The ZPs for 2MASS are the ones in 
Table~\ref{ZPs} here and the ST and AB magnitudes of the Vega SED (which is the same in both papers, hence the advantage of using the same Vega SED throughout the years) for the three 2MASS filters are 
given in Table 4 of \citet{Maiz07a}. Note, however, that the synthetic AB magnitudes calculated in this way are in the default AB system (ZP=0), so they do not correspond to the observed photometric
 magnitudes. An alternative method is given in the appendix of Ma{\'\i}z Apell\'aniz \& Weiler (in preparation).

\end{appendix}

\end{document}